\documentclass[%
 reprint,
 amsmath,amssymb,
 aps,
]{revtex4-2}
\usepackage{amsmath}
\usepackage{amssymb}
\usepackage{braket}
\usepackage{graphicx}
\usepackage{dcolumn}
\usepackage{bm}
\usepackage{hyperref}
\usepackage[mathlines]{lineno}
\hypersetup{colorlinks=true,citecolor=blue,linkcolor=blue,urlcolor=blue}
\usepackage{wasysym}
\UseRawInputEncoding

\begin{document}

\preprint{APS/123-QED}

\title{Observing the relative sign of excited-state dipole transitions \\by combining attosecond streaking and transient absorption spectroscopy}

\author{Shuyuan Hu}
\author{Yu He}%
\author{Gergana D. Borisova}
\author{Maximilian Hartmann}
\author{Paul Birk}
\author{Christian Ott}
\author{Thomas Pfeifer}
\affiliation{%
 Max-Planck-Institut für Kernphysik, Saupfercheckweg 1, 69117 Heidelberg, Germany
}%

\date{\today}

\begin{abstract}
The electronic structure of atomic quantum systems and their dynamical interaction with light is reflected in transition-matrix elements coupling the system's energy eigenstates. In this work, we measure phase shifts of the time-dependent ultrafast absorption to determine the relative signs of the transition-dipole matrix elements. The measurement relies on precise absolute calibration of the relative timing between the used light pulses, which is achieved by combining attosecond transient absorption and attosecond streaking spectroscopy to simultaneously measure the resonant photoabsorption spectra of laser-coupled doubly excited states in helium, together with the attosecond streaked photoelectron spectra. The streaking measurement reveals the absolute attosecond timing and the full temporal profile of the interacting electric fields which is then used to quantify the state-specific dynamics of the measured photoabsorption spectra. By comparing the 1-fs time-scale modulations across the absorption lines corresponding to the $2s2p$ ($^1\textrm{P}$) and $sp_{2,3+}$ ($^1\textrm{P}$) doubly excited states between simulation and measurement, we quantify the signs of the transition dipole matrix elements for the laser-coupled autoionizing states $2s2p$-$2p^2$ and $2p^2$-$sp_{2,3+}$ to be opposite of each other.
\end{abstract}

\maketitle

Light--matter interaction plays an important role in understanding the behavior of atoms and molecules in the presence of electromagnetic radiation. The dipole-matrix elements of laser-coupled transitions establish the connection between the microscopic quantum description and the macroscopic electric field interaction within the electric-dipole approximation. These coupling transition matrix elements play a crucial role in extracting attosecond state-specific dynamics from transient absorption spectra \cite{Argenti_PRA,TDP_Guan,TDP_Yuki,Ott_nature,Viet_PRL,Greene_PRA}. Previous studies that are sensitive to the dipole matrix elements of laser-coupled doubly excited states in helium focused on their magnitude, however, their signs have not been experimentally accessible\cite{Viet_PRL,Ott_nature,Ho_CPL,Loh_CP,Yan_JPB,Aymar_RMP,Mihelič_JPB}. Here we report on an experiment with direct sensitivity to their relative signs by simultaneously measuring the state-resolved transient absorption spectra in combination with attosecond streaking within the same focal volume of a specifically tailored target gas cell.

Attosecond transient absorption spectroscopy (ATAS) has been widely used to access electron dynamics in atoms and molecules on their natural timescale with attosecond resolution \cite{Krausz_nature,Gallmann_MP,Gaarde_JPB}. This technique provides temporal and spectral information of these small quantum systems, which is relevant for fundamental tests of our understanding of ultrafast quantum-state resolved dynamics. Measurements are typically interpreted within the pump-probe concept, where precise relative-timing information of the used pulses --- typically an extreme ultraviolet (XUV) attosecond pulse and a few-cycle near-infrared (NIR) laser pulse --- is crucial for the extraction of the dynamics. Without prior knowledge about the process being studied, it is difficult to extract this information from the experimental data. In attosecond transient absorption spectroscopy the direct measurement of the absolute timing is not straightforward, therefore, previous methods have made use of understanding laser-driven effects that modify the absorption spectrum within the same experiment \cite{Herrmann_NJP}. On the other hand, attosecond streaking spectroscopy is a well-established method with direct access to the full temporal profile of the interacting electric fields and their relative timing \cite{Corkum_PRL_asStreakCamera,Krausz_nature_streaking,Goulielmakis_Science2004,Corkum_streaking}. The streaking spectrograms can thus be used to calibrate the absolute time delay for attosecond transient absorption spectroscopy data. Experiments that combine both methods thus far have either been performed with two different gas target cells in successive streaking and absorption measurements \cite{Wirth_chemPhys}, combining a streaking gas target with absorption in a solid-state sample \cite{Lucchini_science,Volkov_natureP}, or in a two-foci geometry \cite{Lucchini_RevSciInstr}.

In this Letter, we simultaneously measure the resonant photoabsorption spectra of laser-coupled doubly excited states in helium along with the streaked photoelectron spectra within the same target gas and within the same focal volume. We resolve attosecond dynamics across a two-electron wave packet in helium \cite{Ott_nature}, explained by resonantly enhanced two-photon NIR coupling between the lowest doubly excited states of the $sp_{2,n+}$ series, which leads to attosecond time-delay-dependent changes of the measured absorption profiles. In combination with the absolute time delay zero retrieved from an in-situ photoelectron streaking measurement and comparing the absorption spectra with a few-level simulation including the NIR coupling of the relevant states, we obtain an absolute time-delay calibrated absorption spectrogram. This absolute temporal calibration enables direct access to the sign of the product of a closed loop of transition dipole matrix elements, for the example shown here $1s^2 \rightarrow 2s2p \rightarrow 2p^2 \rightarrow  sp_{2,3+} \rightarrow 1s^2$. With the convention of the same sign for the transitions involving the ground state, we are sensitive to the relative signs of the transition dipole matrix elements $\langle 2s2p|\hat{d}|2p^2\rangle$ and $\langle 2p^2|\hat{d}|sp_{2,3+}\rangle$.

\begin{figure*}
\includegraphics[scale=1]{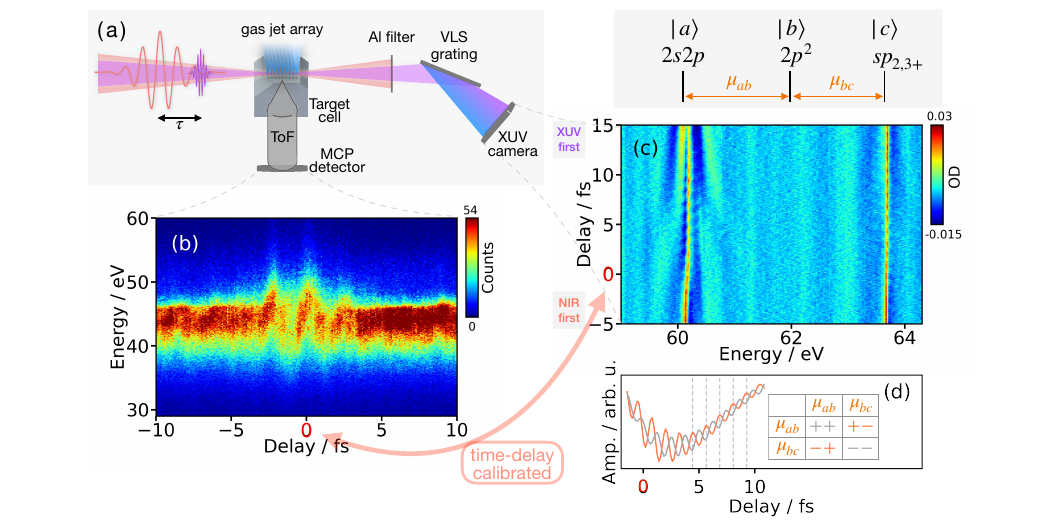}
\caption{\label{fig:1}Experimental scheme and measurements. (a) Experimental setup. (b) Measured streaking trace (1 s integration time per time-delay bin). (c) Measured absorption spectra (50 ms integration time per time-delay bin). The relevant states and dipole matrix elements $\mu_{ab}$ and $\mu_{bc}$ are indicated on top. Time delay ranges from -5 to +15 fs in steps of 85 as. (d) Amplitude of the simulated delay-dependent lineouts of the $2s2p$ state. Orange and grey lines have $\pi$ phase difference, originating from different relative signs for dipole matrix elements $\mu_{ab}$ and $\mu_{bc}$ as illustrated in the inset table.}
\end{figure*}

We simultaneously measure absorption spectra (transmitted photons) and streaking spectra (emitted electrons) as a function of time delay (see Fig.\ref{fig:1}(a)). Details of the experimental setup for transient absorption spectroscopy are described in \cite{Viet_beamline}. In brief, few-cycle carrier-envelope-phase-(CEP)-stable NIR pulses of $\sim$5 fs full width at half maximum (FWHM) duration with 1.54 eV central photon energy (parameters obtained from the streaking retrieval) are used to generate attosecond extreme ultraviolet pulses through high-harmonic generation in neon, at 100 mbar backing pressure. The XUV and NIR pulses are temporally and spatially separated by an interferometric split-and-delay unit and then refocused onto helium gas provided by a specifically designed target cell (Fig.\ref{fig:1}(a)). The target requirements for absorption and streaking are different. For absorption spectroscopy, usually tens of millibars of backing pressure are used inside a static cell, leading to a vacuum chamber pressure of about $10^{-3}$ mbar under steady-state conditions due to gas leaking out through the laser entrance and exit holes. On the other hand, the microchannel plate (MCP) detector used for the photoelectron streaking measurement has to be operated at a pressure below 8$\times$$10^{-6}$ mbar \cite{Kasedorf}. Here, the interaction gas is typically injected by a gas nozzle. Thus, to meet the requirements of both measurements simultaneously and within the same focal volume, one needs to increase the pathlength-density product while maintaining a low chamber pressure. We realize this requirement by a specially designed target gas nozzle which consists of nine evenly spaced holes ($\diameter$=50 $\mathrm{\mu}$m, 200 ${\mathrm{\mu}}$m separation) on the top, forming a gas jet array along the laser propagation direction. This nozzle geometry provides an increased interaction path length for the transient absorption measurements in the forward direction as compared to a single gas jet, while at the same time allows the released electrons to leave the interaction area and not to be blocked by the cell itself. After spectral filtering with an aluminum filter (thickness 200 nm), the XUV spectrum ranges from $\sim$50-72 eV photon energy, and its transmission spectra are measured by a grating-based spectrometer. The streaking detector aligned along the XUV and NIR polarization direction and positioned perpendicular to the laser propagation direction, is a 50-cm-long field-free time-of-flight (ToF) spectrometer from Stefan Kaesdorf \cite{Kasedorf}, adapted with a home-built ToF entrance aperture ($\diameter$=1 mm). To not exceed the vacuum operation limit of the microchannel plate at the back of the ToF tube, and at the same time to acquire as many counts as possible for absorption, the backing pressure of the helium gas target behind the nozzle array is set to 250 mbar.  

\begin{figure}
\includegraphics[scale=1]{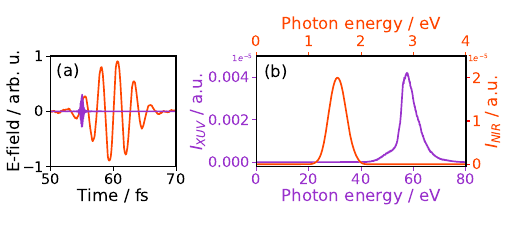}
\caption{The XUV (purple) and NIR (red) fields retrieved from streaking and used in the simulation. (a) Electric fields in the time domain, their position on the time axis is set to a time delay of 5 fs. (b) Intensity spectra of both pulses from the Fourier transform of the fields in the time domain. \label{fig:2}}
\end{figure}
The XUV absorbance of helium in units of optical density (OD) is calculated via the Beer-Lambert law, with its reference spectrum reconstructed using a low-pass Fourier filter, described in detail in \cite{Viet_beamline}. Fig.\ref{fig:1}(c) depicts the resonant absorption $2s2p$ and $sp_{2,3+}$ lines. The coherent NIR coupling between the two states leads to time-delay-dependent modulations on their absorption lines, which evidences a two-electron wave packet \cite{Ott_nature}. Regarding the streaking measurement, the attosecond XUV pulse ionizes helium atoms in the presence of the NIR laser field (streaking field) and the energy spectrum of the photoelectrons released in the process is recorded by the ToF spectrometer. When the streaking field overlaps with the electron wave packet, the time-dependent NIR vector potential is imprinted on the photoelectron spectrum. By measuring the streaked photoelectron spectrum for different time delays (see Fig.\ref{fig:1}(b), we can characterize the XUV pulse. For streaking analysis, we used the extended ptychographic iterative engine (ePIE) retrieval algorithm \cite{Lucchini_ePIE} to reconstruct the exact temporal profile of the electric fields and absolute time delay zero. The retrieved XUV field has (195 $\pm$ 11) as FWHM duration, where the error is obtained from 5 separate runs, and is centered at 54 eV photon energy. The retrieved NIR field has a Gaussian intensity profile with $\sim$5 fs FWHM pulse duration and central photon energy of 1.54 eV (800 nm). We incorporate these retrieved XUV and NIR pulses, plotted in Fig.\ref{fig:2}, into a numerical few-level simulation to directly compare with the measured photoabsorption spectrogram. 

\begin{figure}
\includegraphics[scale=1]{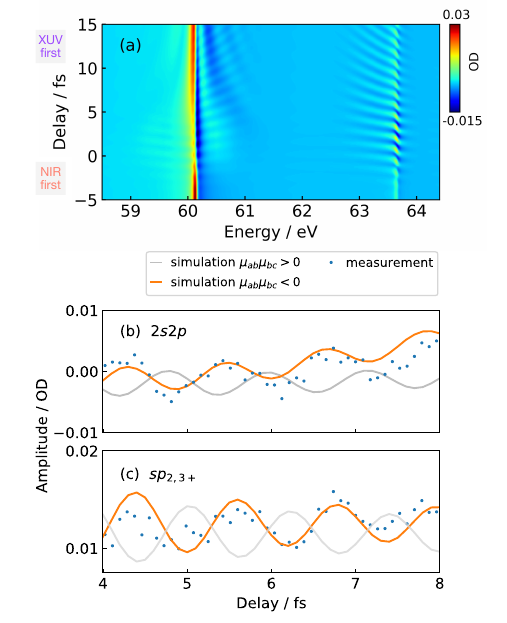}
\caption{Simulated absorption spectrum and energy-integrated lineouts from simulated and measured absorption spectrum. (a) Simulated absorption spectra. (b), (c) Amplitude of the delay-dependent lineout of the $2s2p$ and $sp_{2,3+}$ resonant states. The simulated lines represent the product of the transition dipole moments to be negative ($\mu_{a\mathrm{b}}\mu_{\mathrm{ba}}{<}0$, orange) or positive ($\mu_{\mathrm{ab}}\mu_{\mathrm{ba}}{>}0$, grey). The case $\mu_{\mathrm{ab}}\mu_{\mathrm{ba}}{<}0$ shows good agreement with the delay-calibrated measured data.\label{fig:3}}
\end{figure}

The simulation employs a few-level model based on the NIR coupling of the relevant discrete states via the time-dependent Schrödinger equation. In the model, the system consists of the ground state $\ket{g}$, four doubly excited bound states $2s^2$ ($^1\textrm{S}^e$), $2s2p$ ($^1\textrm{P}^o$), $2p^2$ ($^1\textrm{S}^e$), and $sp_{2,3+}$ ($^1\textrm{P}^o$), and two continua $\ket{1s,\varepsilon p}$ and $\ket{1s,\varepsilon s}$, into which the $^1\textrm{P}^o$ and $^1\textrm{S}^e$ bound states autoionize, respectively. Both continua are approximated each by a short-lived state of $\sim$21 eV width, effectively modelling ionization loss. The states $2s^2$ and $2p^2$ are included because they are in close NIR resonance with the $2s2p$ state. Other states from the $N{=}2$ doubly excited Rydberg series are far off-resonance with respect to the coupling NIR laser or are significantly lower in coupling strength compared to the states included in this model, and thus are neglected. The $2s2p$ and $sp_{2,3+}$ states are excited by the XUV pulse from the ground state, and non-perturbatively coupled by the time-delayed NIR pulse. The total Hamiltonian is $H{=}H_0+H_{\textrm{NIR}}+H_{\textrm{XUV}}+H_{\textrm{CI}}$, where $H_{0}$ is the field-free Hamiltonian and the subsequent terms describe the interaction of the atom with the NIR laser ($H_{\textrm{NIR}}$), the attosecond XUV pulse ($H_{\textrm{XUV}}$), and the configuration interaction ($H_{\textrm{CI}}$) of each discrete state with its respective continuum state, respectively. This model allows to calculate the attosecond time-resolved XUV absorption spectrum, which can be directly compared to the measured spectra. In the simulation we use a numerical step size of 0.006 fs and total simulation time of 774 fs, allowing sufficient time for all states to decay. The simulation parameters are tuned to reproduce the energy position, decay width and asymmetry of the XUV transition from the ground state to the autoionizing $2s2p$ and $sp_{2,3+}$ doubly excited states \cite{Rost_JPB}. The absolute values of the dipole matrix elements regarding near-resonant NIR coupling between the bound states $2s^2$-$2s2p$, $2s2p$-$2p^2$ ($\mu_{\mathrm{ab}}$), and $2p^2$-$sp_{2,3+}$ ($\mu_{\mathrm{bc}}$) are calculated by theory as 1.52 a.u., 2.16 a.u, and 0.81 a.u. \cite{Mihelič_JPB,Ott_nature}, respectively. We estimate the NIR peak intensity on the order of 0.7 TW/cm$^2$ by comparing the measured and simulated absorption spectra. The simulated absorption spectra are shown in Fig.\ref{fig:3}(a). 

When the XUV pulse arrives before or during the NIR-pulse duration, the absorption is modulated as a function of delay with half-NIR-cycle (2$\omega_{NIR}$) oscillations, resulting from a phase-coherent excitation of the $2s2p$ and $sp_{2,3+}$ states that are resonantly coupled by two NIR photons through the $2p^2$ state. From the time-delay-calibrated simulated absorption spectrum (see Fig.\ref{fig:3}(a)), we can extract the delay-dependent lineouts at the $2s2p$ and $sp_{2,3+}$ resonances, and directly compare with the corresponding absorption lineouts of the measurement, which is shown in Fig.\ref{fig:3}(b). Because the modified asymmetric lineshape of the resonance introduces a phase change across the resonant state \cite{Kaldun_PRL}, it is important to integrate only over a small energy region when comparing the lineouts. For the comparison between experiment and theory, an energy region of $\sim$0.045 eV was chosen, for $2s2p$ the integration limits are 60.135--60.180 eV, and for $sp_{2,3+}$ 63.645--63.690 eV. Different relative signs of $\mu_{\mathrm{ab}}$ and $\mu_{\mathrm{bc}}$ lead to a $\pi$ phase shift in the oscillation of delay-dependent lineouts as illustrated in Fig.\ref{fig:1}(d). As can be seen in Fig.\ref{fig:3}(b), we find that the measurement agrees well with the attosecond-calibrated simulation when the relative signs of the transition dipole matrix elements $\mu_{ab}$ and $\mu_{bc}$ are opposite of each other. Simulations with different signs of the dipole matrix element for the coupling of the $2s2p$ and $2s^2$ states have also been performed, which show no influence on the resonance line shape of the $2s2p$ state. With the in-situ streaking measurement within the same focal volume, we are thus able to obtain a time-delay zero calibrated absorption measurement on the attosecond timescale, and thus to determine the relative signs of the involved transitions.   

In conclusion, with our simultaneous measurement, we can retrieve the absolute timings from attosecond streaking and use this information to obtain a time-delay zero calibrated absorption measurement in situ, and in the same focal volume. With the help of a few-level simulation model of the relevant NIR-laser-coupled states, we are able to compare the delay-dependent energy-integrated lineouts of the resonantly excited states with the experiment. As a key result, we quantify the relative signs of the dipole matrix elements $\mu_{ab}$ and $\mu_{bc}$ to be opposite of each other. With this novel experimental method, one can make full use of the state-resolved spectral and attosecond temporal information. The extraction of information from the spectral lineshapes combined with time-dependent phase and amplitude modifications induced by resonant laser coupling and an absolute time-zero reference, opens the possibility for a new class of state-resolved attosecond experiments and the underlying control of electron dynamics.


\bibliography{bibliography}

\end{document}